\documentclass[12pt,preprint]{aastex}

\newcommand{\etal}{et~al.}
\newcommand{\fig}{Figure~}
\newcommand{\mic}{\,$\mu$m}
\newcommand{\av}{A_V}
\newcommand{\water}{H$_2$O}
\newcommand{\cod}{CO$_2$}

\shorttitle{Uptake of interstellar gaseous CO into ices}
\shortauthors{Whittet et al.}

\slugcomment{Version date: \today}

\begin{document}

\title{THE UPTAKE OF INTERSTELLAR GASEOUS CO INTO ICY GRAIN MANTLES IN A QUIESCENT DARK CLOUD\\ ~}
		
\author{D.~C.~B. Whittet}

\affil{New York Center for Astrobiology, and Department of Physics, Applied Physics \& Astronomy,
	Rensselaer Polytechnic Institute, 110 Eighth Street, Troy, NY 12180, USA.}

\author{P.~F. Goldsmith and J.~L. Pineda}

\affil{Jet Propulsion Laboratory, California Institute of Technology, 4800 Oak Grove Drive, Pasadena, CA 91109, USA.\\ ~}

\begin{abstract}
Data from the Five College Radio Astronomy Observatory CO Mapping Survey of the Taurus molecular cloud are combined with extinction data for a sample of 292~background field stars to investigate the uptake of CO from the gas to icy grain mantles on dust within the cloud. On the assumption that the reservoir of CO in the ices is well represented by the combined abundances of solid CO and solid \cod\ (which forms by oxidation of CO on the dust), we find that the total column density (gas + solid) correlates tightly with visual extinction ($\av$) over the range $5<\av < 30$~mag, i.e., up to the highest extinctions covered by our sample. The mean depletion of gas-phase CO, expressed as $\delta({\rm CO}) = N({\rm CO})_{\rm ice}/N({\rm CO})_{\rm total}$, increases monotonically from negligible levels for $\av\la 5$ to $\sim$\,0.3 at $\av = 10$ and $\sim$\,0.6 at $\av = 30$. As these results refer to line-of-sight averages, they must be considered lower limits to the actual depletion at loci deep within the cloud, which may approach unity. We show that it is plausible for such high levels of depletion to be reached in dense cores on timescales $\sim 0.6$~Myr, comparable with their expected lifetimes. Dispersal of cores during star formation may be effective in maintaining observable levels of gaseous CO on the longer timescales estimated for the age of the cloud.
\end{abstract}



\section{Introduction}
\label{intro}
CO is a vital molecule in interstellar astrophysics for several reasons. For example, CO has long been used as a proxy for H$_2$, allowing the distribution of molecular gas to be mapped by means of the readily-observable gas-phase emission lines at millimeter wavelengths. As a repository for substantial fractions of the available elemental carbon and oxygen, CO also plays an important role in the chemical evolution of the interstellar medium (ISM) as a reactant in processes that may lead to synthesis of complex molecules. An important issue that affects both of these areas of research is the degree to which CO condenses out of the gas onto dust grains in molecular clouds. From a theoretical standpoint, it is expected that CO will stick efficiently to grain surfaces at the low temperatures ($T\la 17$~K) prevailing in dense regions remote from embedded stars, suggesting that virtually all of the CO could be removed from the gas on timescales shorter than cloud lifetimes (L\'eger 1983; Walmsley \etal\ 2004) unless there is an efficient desorption mechanism (Roberts \etal\ 2007; {\"O}berg \etal\ 2007). The presence of solid CO has indeed been inferred from density-correlated depletions in the observed gas-phase abundance toward cloud cores (e.g., Bacmann \etal\ 2002), and detected directly in some lines of sight by means of its 4.67\mic\ absorption feature in the infrared spectra of stars observed through molecular-cloud material (e.g., Chiar \etal\ 1995; Pontoppidan \etal\ 2003).

To quantify the distribution of CO between gaseous and solid phases reliably as a function of physical conditions is thus an important astrophysical goal. It may be accomplished by measuring and comparing gas-phase and solid-state CO column densities in the same lines of sight. However, this apparently simple task proves to be quite difficult to accomplish in practice. CO-bearing ice mantles that form on the dust are detectable only in absorption, by means of their vibrational transitions in the infrared, so observations of the solid phase are restricted to lines of sight toward background continuum sources, typically young stellar objects (YSOs) embedded within molecular clouds, or field stars located behind them. In principle, gas-phase CO may be studied in either absorption or emission, via its electronic transitions, vibration-rotation bands or pure rotation features at ultraviolet, infrared and millimeter wavelengths, respectively. The most reliable method would be to study solid and gaseous phases in absorption simultaneously, in the same lines of sight, against the same continuum sources. Acquisition of the necessary data is relatively straight-forward for CO$_{\rm ice}$, far more challenging for CO$_{\rm gas}$. Studies of the ultraviolet absorption features of gaseous CO (e.g., Federman \etal\ 1994, 2003) are limited to lines of sight with extinctions ($\av\la2$) well below the level at which ice is detected. The gas-phase lines within the infrared vibration-rotation bands are narrow and suffer from significant telluric contamination in ground-based observations, requiring both high spectral resolving power ($R\ga 10^4$) and excellent signal-to-noise for effective observation. Data of sufficient quality have been acquired to date only for a small number of relatively bright sources (Mitchell \etal\ 1988, 1990; Shuping \etal\ 2001; Rettig \etal\ 2005), and these turn out to be exclusively YSOs of high or intermediate mass. The further possibility to observe pure rotational transitions in absorption proves to be infeasible because of the lack of sufficiently strong background sources at millimeter wavelengths and the need to observe multiple transitions to account for the total column density, which is spread over many $J$-states. 

In summary, to date, no convincing detection of interstellar gaseous CO has has been made in absorption toward any field star located behind a significant column of dense molecular-cloud material, which is unfortunate, as it is the field stars that yield the most reliable information on CO depletion in quiescent regions of the clouds, whereas embedded YSOs may drive sublimation of ices in their local environment. To measure CO depletion in quiescent clouds it is necessary to estimate gas phase column densities from millimeter-wave observations of CO {\it emission\/} in lines of sight toward background field stars, for comparison with the infrared absorption-line data for solid CO. This raises an obvious concern that the two types of observation might not sample identical regions of space, because of differences in effective beam size and the possibility of material behind the star, which contributes to the millimeter-wave intensity but not to the ice absorption. The well-known problem of saturation in the gas-phase CO emission lines is a further concern, generally necessitating observation of rarer isotopologues such as $^{13}$CO, C$^{18}$O, C$^{17}$O and $^{13}{\rm C}^{18}$O (e.g., Frerking \etal\ 1982; Bensch \etal\ 2001a; Harjunp{\"a}{\"a} \etal\ 2004).

This paper compares gas-phase CO column densities extracted from the Five College Radio Astronomy Observatory (FCRAO) CO Mapping Survey of the Taurus molecular cloud (Goldsmith \etal\ 2008; Narayanan \etal\ 2008; see Section~2) with extinction data for some 292~background field stars in the same Galactic region (Shenoy \etal\ 2008; Whittet \etal\ 2001). The spatial resolution of the gas-phase CO observations from the FCRAO survey ($\sim 45^{\prime\prime}$) is improved by a factor $\sim 2$ compared with an earlier study by Frerking \etal\ (1982) for a sample of 14~background stars in the Taurus region. Contamination of the gas-phase column densities by molecular material behind the stellar sources in our sample is unlikely to be a significant problem, given the location of the cloud away from the Galactic equator ($b\sim 15^\circ$) in a direction where virtually all of the observed interstellar material is confined to the cloud itself. Column densities of the ices have been shown in previous work on a subset of our sample to be well correlated with $\av$ (Whittet \etal\ 2007 and references therein), enabling reasonable estimates to be made in cases where direct measurements are unavailable. Gas-phase and solid-state abundances may thus be compared over the entire sample. We use our results to show (Section~3) that the mean line-of-sight depletion of CO from gas onto dust increases systematically from negligible levels at low extinction to $\sim 60$\% at $\av\sim 30$. Our results are in broad agreement with theoretical predictions based on expected timescales for the depletion of gaseous CO in quiescent cloud cores (Section~4). 

\section{Gas-phase CO column density determination}
\label{nco}
We use the $^{12}$CO and $^{13}$CO observations of the Taurus molecular cloud presented by Goldsmith \etal\ (2008) to obtain gas-phase
column densities, $N$(CO)$_{\rm gas}$, at the positions of our program stars. The primary sample is a catalog of 247~reddened background field stars identified from Two-Micron All Sky Survey (2MASS) photometry and other infrared data by Shenoy \etal\ (2008, their Table~1), covering the extinction range $3<\av<29$. To enable better coverage of low extinctions, we add optically-selected field stars with well-determined extinction values from Whittet \etal\ (2001) and references therein (see \fig 5 of Shenoy \etal\ 2008 for a histogram of the distributions), thus increasing the sample size to 292. Finally, we include for comparison ten highly reddened YSOs from the Shenoy \etal\ study (their Table~2). All stellar positions used in the current work are taken from the 2MASS Point Source Catalog (Skrutskie \etal\ 2006). The complete sample is listed in Table~1 (online version).

We first calculated the column density of $^{13}$CO assuming that its $J = 1
\to 0$ transition is optically thin, that $^{12}$CO $J = 1 \to 0$ is
optically thick, and that local thermodynamic equilibrium applies.
The column density of $^{13}$CO is then proportional to its
integrated intensity and inversely to the spontaneous decay rate.  It also
depends on the excitation temperature, derived from $^{12}$CO,
which affects the upper level population and the partition function.
We also determine the optical depth, which is a function of the
ratio between the $^{13}$CO intensity and the excitation
temperature, to apply a correction for saturation.  Finally, we
transform the column density of $^{13}$CO to that of $^{12}$CO
assuming an isotopic ratio of 65 (Bensch \etal\ 2001a) in the relatively well-shielded portions
of the cloud defined by the ``mask~2" region of Goldsmith \etal\ (2008). 

The determination of $N$(CO)$_{\rm gas}$ was improved relative to that
presented by Goldsmith \etal\ (2008) by including an updated value of
the spontaneous decay rate, using an exact numerical rather than approximate
analytical calculation of the partition function. Additionally, the data
were corrected for error beam pick-up using the method presented by Bensch 
\etal\ (2001b). The resulting values of $N$(CO)$_{\rm gas}$ are about
$\sim$20\% larger than those of Goldsmith \etal\ (2008). Full
details of these improvements are presented elsewhere (Pineda \etal\ 2010).

Motivated by observations of core-to-edge temperature differences in
molecular clouds (Evans \etal\ 2001) which can be found even in regions of only
moderate radiation field intensity, Pineda \etal\ (2010) studied the effects of
temperature gradients on the determination of $N$(CO)$_{\rm gas}$. The radiative transfer
code RATRAN (Hogerheijde \& van der Tak 2000) was used to model the $^{12}$CO and $^{13}$CO
emission emerging from a model cloud. Pineda \etal\ found that using $^{12}$CO
to determine the excitation temperature of the CO gas only traces the
temperature at low column densities while the excitation temperature is overestimated
for larger column densities.  This produces an underestimate of the $^{13}$CO optical depth, and in consequence 
the opacity correction of $N(^{13}{\rm CO})$, which is usually evaluated assuming an isothermal cloud. The column densities
presented here were corrected by this method including modest edge-center temperature gradients of $\sim4$~K.
This procedure typically increases the value of $N$(CO)$_{\rm gas}$ by $\sim$10--40\%, with larger increases
occurring at higher column densities. Final values are presented in Table~1, together with ice and extinction data.

\section{Results and discussion}
\subsection{Gaseous CO versus extinction}
A plot of $N({\rm CO})_{\rm gas}$ against $\av$ is shown in \fig 1. In addition to results from the current work, we include $N({\rm CO})_{\rm gas}$ estimates from the data of Frerking \etal\ (1982) for lines of sight to 14~Taurus field stars\footnote{The abundance ratios we used to estimate $N({\rm CO})_{\rm gas}$ in the common isotopic form from the various isotopologues observed by Frerking \etal\ (1982) were $^{12}$CO/$^{13}$CO = 65, C$^{16}$O/C$^{18}$O = 560, and C$^{16}$O/C$^{17}$O = 1790 (Bensch \etal\ 2001a; Ladd 2004; Wilson \& Rood 1994).} in combination with $\av$ values from Shenoy \etal\ (2008). Previous observations of gas-phase C$^{18}$O toward several molecular clouds suggest a general linear relation 
\begin{equation}
N({\rm C^{18}O})_{\rm gas} = 2 \times 10^{14} (A_V-2)~~{\rm cm}^{-2}
\end{equation}
for $\av$ values up to about 10~mag (see Kainulainen \etal\ 2006 and references therein), and this relation is represented (after conversion to the common isotopic form) by a dotted line in \fig 1. The distribution of the data is, indeed, consistent with this relation at low extinction. However, a clear divergence is apparent for $\av> 10$, with little further increase in $N({\rm CO})_{\rm gas}$ with $\av$. The general trend is well represented by a sigmoidal fit\footnote{The functional form is $y = a_2 + (a_1-a_2)/(1 + (x/x_0)^p)$; see http://www.originlab.com. Note that the fit is purely empirical, for illustrative purposes, and has no basis in theory.} shown as a solid curve in \fig 1. It is possible that this behavior might result in part from residual saturation in our column density calculations (Section~2), leading to underestimates at the highest optical depths; however, similar behavior is seen in the data of Frerking \etal\ (1982) for rarer isotopologues such as C$^{17}$O and $^{13}$C$^{18}$O that are presumed to be unsaturated (\fig 1), and we therefore consider residual saturation in our data to be at worst a second order effect. Note also that a similar trend in $N$(CO)$_{\rm gas}$ with $\av$ was found in another quiescent dense cloud (IC~5146) by Kramer \etal\ (1999), based on observations of C$^{18}$O.

\subsection{Adding the solid phase}
Previous observations have shown that both CO and \cod\ are present in the ices over the same range of $\av$ in which gas-phase CO appears to become depleted in the Taurus cloud (Whittet \etal\ 1989, 1998, 2007; Chiar \etal\ 1995; Bergin \etal\ 2005; Knez \etal\ 2005). Astrochemical models and laboratory simulations predict that \cod\ forms in situ on interstellar grains, as the result of simple surface oxidation reactions such as \hbox{CO + O $\rightarrow$ \cod} and \hbox{CO + OH $\rightarrow$ \cod\ + H} (Tielens \& Hagen 1982; Ruffle \& Herbst 2001; Roser \etal\ 2001). A substantial fraction of the CO that becomes depleted onto dust is thus predicted to be converted to \cod\ on timescales short compared with cloud lifetimes, and so a complete inventory of adsorbed CO should include the reservoir represented by solid \cod\ as well as by CO itself\footnote{Note that \cod\ has negligible abundance in the gas-phase in cold molecular clouds: interstellar gaseous \cod\ is observed only in lines of sight to luminous YSOs that drive rapid sublimation of the ices (van Dishoeck \etal\ 1996).}. An estimate of the ``total CO" column density that accounts for both depleted CO and that remaining in the gas is therefore given by
\begin{equation}
N({\rm CO})_{\rm total} = N({\rm CO})_{\rm gas} + N({\rm CO})_{\rm ice} + N({\rm CO}_2)_{\rm ice}.
\end{equation}
No other CO-bearing molecule appears to be sufficiently abundant in quiescent clouds to constitute a significant reservoir of CO in the solid phase: methanol (CH$_3$OH), for example, is sometimes abundant toward YSOs but appears to be no more than a minor constituent of the ices observed toward field stars (Chiar \etal\ 1996; Gibb \etal\ 2004). 

The observations also show that each of the major ice constituents (\water, CO and \cod) displays a tight linear correlation between column density and extinction. These correlations may be expressed in the form
\begin{equation}
N_{\rm \,ice} = q(\av - \av^{~0})
\end{equation}
where $\av^{~0}$ is the ``threshold" extinction for detection of a given species, such that $N_{\rm ice} = 0$ for $\av\le\av^{~0}$. The best available values of the constants $q$ and $\av^{~0}$ for the two species relevant to the present study are:
\begin{eqnarray}
 {\rm  CO}:~~~q=(0.400\pm 0.060) \times 10^{17}\,{\rm cm^{-2}\,mag^{-1}},~A_V^{~0}=6.7\pm 1.6~{\rm mag}~\\
 {\rm CO}_2:~~q=(0.252\pm 0.036) \times 10^{17}\,{\rm cm^{-2}\,mag^{-1}},~A_V^{~0}=4.3\pm 1.0~{\rm mag}.
\end{eqnarray}
These values are based on fits to data for a sample of 13~Taurus field stars in the extinction range $5<\av<24$~mag (see, e.g, Figure~4 of Whittet \etal\ 2007). The majority of stars in our sample lack the spectral data that would enable direct measurement of ice column densities: in these cases, we use Equation~(3) with the above values of $q$ and $\av^{~0}$ to estimate the contributions of solid CO and \cod\ to the total column density in Equation~(2) from the $\av$ values for field stars. Resulting values are listed in Table~1. We consider the well-established $N_{\rm \,ice}$ versus $\av$ correlations to be sufficiently strong to justify this approach for field stars, but it may not be appropriate for YSOs. For this reason, total CO column densities are calculated for YSOs only in cases where direct measurements of both $N({\rm CO})_{\rm ice}$ and $N({\rm CO}_2)_{\rm ice}$ are available from the literature: only two out of ten in our sample (J04395574+2545020 and J04400800+2605253) satisfy this requirement (Cook \etal\ 2010). 

The resulting plot of $N({\rm CO})_{\rm total}$ versus $\av$ is shown in \fig 2. The overall distribution of points is clearly consistent with the linear least-squares fit (solid line), yielding a correlation only marginally different from that predicted by Equation~(1). This corresponds to an abundance $N_{\rm CO}/N_{\rm H} \sim 5.3\times 10^{-5}$ at extinctions above the threshold, assuming the canonical value of $N_{\rm H}/\av \approx 1.9\times 10^{21}~{\rm cm^{-2}\,mag^{-1}}$ for the total hydrogen gas to extinction ratio (Bohlin \etal\ 1978).

\subsection{The CO depletion factor}
The CO depletion factor is defined
\begin{equation} 
\delta({\rm CO}) = N({\rm CO})_{\rm ice}/N({\rm CO})_{\rm total} = 1 - N({\rm CO})_{\rm gas}/N({\rm CO})_{\rm total},
\end{equation}
such that $0\le\delta({\rm CO})\le 1$, the lower and upper bounds corresponding to all CO in gaseous and solid forms, respectively. This quantity was calculated for each line of sight in our sample from the data used to construct Figures~1 and 2, and the results plotted against $\av$ in \fig 3. The curve is calculated in the same way by ratioing the sigmoidal fit from \fig 1 with the linear fit from \fig 2. In general, the data in \fig 3 are consistent with a monotonic increase in $\delta({\rm CO})$ from zero at low extinction ($\av \la 5$) to $\sim 0.6$ at the high end of the observed range. The scatter appears somewhat asymmetric because it arises primarily in the denominator of Equation~(6); outliers well above the trend arise because a few lines of sight with intermediate extinction have unexpectedly low gaseous CO column densities for their $\av$ values.

It is important to bear in mind that $\delta({\rm CO})$ measures the {\it mean\/} depletion along the line of sight, and this may differ from the local value at a given location corresponding to a given optical depth. In particular, the actual depletion at high $\av$ deep inside the cloud is expected to exceed the value of $\sim0.6$ suggested by \fig 3, because of the ``dilution" effect of lower-density regions that are inevitably included in any line of sight passing through the cloud. To illustrate this, we compare data in Table~2 for typical lines of sight at $\av = 5$, 10 and 20 (average values) and at $\av = 28.7$ (for the individual star J04135352+2813056, which has the highest extinction in our sample). The final column lists differential depletion factors, calculated by replacing total line-of-sight column densities in Equation~(6) with incremental values relative to the previous row: for example, the value $\delta({\rm CO})=0.69$ represents a segment of column through the cloud extending from $\av = 10$ to $\av = 20$. Comparing mean and differential values of $\delta({\rm CO})$ in Table~2, we find that the latter increase more rapidly with $\av$, and approach 0.9 (i.e., 90\% depletion from the gas) at the highest extinctions sampled by our data.

Our results show that CO depletion is not only high within dense cores but is also substantial at intermediate levels of extinction. This is contrary to what is often assumed. For example, Padoan \etal\ (2006) report significant differences in the distributions of gas and dust in the Taurus cloud based on a comparison of maps of gas-phase CO and extinction, which they attribute to spatial fluctuations in the gas-to-dust ratio driven by turbulent flows. These authors assume in their analysis that the CO remains entirely in the gas for $\av < 10$~mag; however, our results indicate that the depletion is already significant ($\sim$\,30\% or more) at this level of extinction. 

\subsection{Comparison of field stars and YSOs}
The ten young stellar objects in our sample display an unremarkable distribution in Figure~1, having gaseous CO column densities similar to those of field stars in the same extinction range. For reasons discussed above, only two YSOs are included in the calculation of total CO column density and depletion (Figures~2 and 3). These objects have estimated mean depletion factors of 0.36 (J04400800+2605253) and 0.47 (J04395574+2545020), values that are again consistent with those of field stars in the same extinction range. This is in contrast to the results of Mitchell \etal\ (1990), who find $\delta({\rm CO}) \la 0.1$ for a sample of seven YSOs. The most likely explanation is a systematic difference in mass between the two samples: massive YSOs such as those in the Mitchell \etal\ study evolve rapidly and become sufficiently luminous to drive widespread sublimation of interstellar ices in their vicinity, whereas the lower-mass objects in our sample have less impact on their environment. For a young star of a given mass, the degree of depletion of CO and other molecules in the envelope is likely to evolve with time (e.g., Thomas \& Fuller 2008). Rettig \etal\ (2005) estimate that the solid CO abundance exceeds the gas phase abundance by a factor $\sim6$ toward the eruptive T~Tauri star V1647 Orionis, which implies $\delta({\rm CO})\ga 0.85$ in this line of sight. It would clearly be important to obtain data for a larger sample of low and intermediate-mass YSOs. 

In the case of one object in our sample, the intermediate-mass young star Elias~18 (J04395574+2545020), a notable discrepancy is apparent between our value of $N({\rm CO})_{\rm gas} \approx 1.0 \times 10^{18}~{\rm cm^{-2}}$ from the radio survey data and that of Shuping \etal\ (2001), who obtain a value a factor $\sim100$ less from a study of the infrared spectrum of this source. It is possible that the difference might arise because of the finite beam size of the radio observations, and the possibility of dense material behind the source, as previously discussed in Section~1. Shuping \etal\ propose that CO is highly depleted in the circumstellar disk of Elias~18 and that it suffers very little foreground absorption in the molecular cloud; however, we consider this unlikely, given that the abundance of solid CO is actually less toward Elias~18 compared with the Taurus field stars with respect to both \water-ice and total extinction ($\av\approx 22$~mag) in the line of sight (Chiar \etal\ 1995; Nummelin \etal\ 2001). Hence, the {\it total\/} CO as well as the gas-phase CO toward Elias~18 must be unusually low if the Shupping \etal\ result is accurate.

\section{Comparison with models}
In this section we briefly review and discuss theoretical constraints on the timescale for CO depletion in molecular clouds, and examine the degree to which they are consistent with our results. The rate at which gaseous atoms or molecules accumulate onto the dust depends on the collision rate and the probability that a collision will lead to attachment. The collision rate depends on the thermal speed of the gaseous species and the surface area of the dust. The latter may be expressed in terms of the total grain geometric cross-sectional area per H atom:
\begin{equation}
\Sigma_{\rm d} = \int^{a_{\rm max}}_{a_{\rm min}} {n_{\rm d}(a)\,\sigma_{\rm d}(a)\over n_{\rm H}} ~da
\end{equation}
where $a$ represents an appropriate grain dimension such as the radius for a sphere, $\sigma_{\rm d}(a)$ is the cross-sectional area of a grain, $n_{\rm d}(a)$ is the number density of dust in the size range $a \rightarrow a + da$, and $n_{\rm H} = n({\rm H\,I}) + 2n({\rm H_2})$ is the total hydrogen (proton) density (see, e.g., Spitzer 1978; Draine 1985). It is usual to evaluate $\Sigma_{\rm d}$ assuming spherical particles and a power-law size distribution $n_{\rm d}(a)\propto a^{-3.5}$ (the ``MRN" size distribution of Mathis \etal\ 1977), with the upper and lower limits $a_{\rm max}$ and $a_{\rm min}$ set by optimization of grain models fit to the mean interstellar extinction curve for the diffuse ISM. This approach yields $\Sigma_{\rm d} \approx 1.0\times 10^{-21}~{\rm cm}^2~{\rm H}^{-1}$.

Assuming that grain speeds are small compared with the mean thermal speed ($v$) of impinging gas particles, the timescale for depletion is given by 
\begin{equation}
t_{\rm dep} = {1\over n_{\rm H}\,\xi\,\Sigma_{\rm d}\,v}
\end{equation}
where $v=(8kT_{\rm gas}/\pi m)^{1/2}$ for particles of mass $m$, and $\xi$ is the sticking probability. Both theoretical models and experimental data predict a high sticking probability ($\xi\approx 1$) for CO at molecular-cloud temperatures (L\'eger 1983; Bisschop \etal\ 2006). Assuming $T_{\rm gas} = 12$~K (e.g., Stepnik \etal\ 2003) and setting $m$ to the mass of CO, Equation~(8) then yields the result 
\begin{equation}
t_{\rm dep} \sim {3\times 10^9\over n_{\rm H}}~{\rm yr} 
\end{equation}
where $n_{\rm H}$ is in units of cm$^{-3}$. 

A systematic error in estimates of $t_{\rm dep}$ that is generally overlooked arises because the calculation of $\Sigma_{\rm d}$ assumes a grain size distribution appropriate to diffuse (low-density) phases of the ISM. In dense clouds, the grains grow by coagulation as well as by adsorption of atoms and molecules from the gas. This can lead to a substantial reduction in the relative number of small grains, and hence in cross-sectional area, on timescales short compared with cloud lifetimes. Draine (1985) estimates that for spherical grains following the MRN size distribution, 80\% of $\Sigma_{\rm d}$ is contributed by grains with radii $a<0.05$\mic\ that contain only $\sim$\,30\% of the total mass of dust. Coagulation may thus lead to substantial reduction in $\Sigma_{\rm d}$, with proportionate increase in $t_{\rm dep}$. Observational evidence for growth by coagulation is provided by systematic changes in the extinction curve, characterized by the parameter $R_V = A_V/E_{B-V}$ (the ratio of total to selective extinction; Fitzpatrick 1999; Whittet 2003). Consider, for example, ultraviolet extinction at wavelength $\lambda = 0.1$\mic, which is where particles of radius \hbox{$a\sim\lambda/2\pi\sim 0.02$\mic} contribute most efficiently: the extinction is observed to decline by a factor $\sim$\,2 as $R_V$ increases from the average diffuse-ISM value of 3.1 to a typical value of 5.5 within a dense cloud (Whittet \etal\ 2001), implying a corresponding reduction in $n_{\rm d}$ for grains in this size range. Equation~(9) may thus underestimate the depletion timescale within molecular clouds by a factor of order~2. A typical dense core within the Taurus complex has density $n_{\rm H}\sim 10^4~{\rm cm}^{-3}$ (see, e.g., Table~1 of Onishi \etal\ 1996). Allowing for a doubling of the constant in Equation~(9), this density corresponds to a CO depletion timescale $\sim 0.6$~Myr. 

The age of the Taurus complex is at least a few Myr (e.g., Kenyon \& Hartmann 1995). Estimates of this order for Taurus and other dark clouds are often used in the literature in conjunction with variants on Equation~(9) to support the view that all the available CO is expected to be in solid form (and hence to be unobservable at millimeter wavelengths) on timescales much less than cloud lifetimes, unless an efficient (and poorly understood) desorption mechanism is operative (e.g., Roberts \etal\ 2007, {\"O}berg \etal\ 2007 and references therein). For a cloud such as Taurus, in which star formation has been ongoing for at least 1--2~Myr (Kenyon \& Hartmann 1995), the mechanism may simply be star formation itself. Protostars form within prestellar cores and disperse them on timescales $<1$~Myr (Kenyon \& Hartmann 1995; Onishi \etal\ 2002; Visser \etal\ 2002). The life expectancy of an individual core is a factor of 5--10 less than that of the entire complex.

To summarize, in general these estimates seem consistent with our results. It is highly plausible that CO depletion factors exceeding 50\% can be reached within dense cores on timescales comparable with their expected lifetimes. However, no crisis is implied in our understanding of desorption mechanisms by the observation that gas phase CO remains observable over the range of densities covered by our survey.

\section{Conclusion}
We have studied the uptake of CO from gas to dust in the Taurus molecular cloud by comparing CO column densities from millimeter-wave observations with extinction values and ice column densities determined from infrared observations of stars. The main conclusion of this work is that the mean depletion of gas-phase CO increases monotonically from negligible levels at visual extinctions $\av\la 5$ to about 60\% at $\av\sim 30$. The depletion is significant ($\sim$\,30\%) even at moderate levels of extinction ($\av\sim 10$). These results refer to line-of-sight averages and should be considered as lower limits to the actual depletion at loci deep within the cloud, which may approach 90--100\% in dense cores. We estimate the timescale for CO depletion in a core of density $n_{\rm H}\sim 10^4~{\rm cm}^{-3}$ to be $\sim 0.6$~Myr, a value consistent with independent estimates of the lifetime of a typical core. Dispersal of cores during star formation may help to maintain observable levels of gaseous CO over cloud lifetimes. Our results will provide useful constraints on astrochemical models for molecular clouds (e.g., Chang \etal\ 2007; Cuppen \& Herbst 2007), and on estimates of macroscopic cloud properties such as mass, structure, and the distribution of molecular material (e.g., Padoan \etal\ 2006; Pineda \etal\ 2010). 

The constraints placed on CO depletion by our analysis probably represent the limits of what can be accomplished by combining data from the disparate techniques of gas-phase emission spectroscopy and solid-state absorption spectroscopy. Further refinement awaits a targeted survey of infrared vibration-rotation lines in absorption toward background field stars known to display ice features, requiring 4.5--4.8\mic\ spectra of high signal-to-noise with resolving powers $\ga 10^4$. We also advocate further study of intermediate-mass YSOs by this technique, to investigate the degree to which they may contribute to the global recycling of CO between solid and gaseous phases in molecular clouds, and to better evaluate other astrochemical pathways for CO, such as CH$_3$OH formation, in protostellar envelopes.

\acknowledgments
This research has made use of data from the Two Micron All Sky Survey, which is a joint project of the University of Massachusetts and the Infrared Processing and Analysis Center, funded by the National Aeronautics and Space Administration (NASA) and the National Science Foundation. D.C.B.W.\ acknowledges financial support from the NASA Exobiology and Evolutionary Biology program (grant NNX07AK38G) and the NASA Astrobiology Institute (grant NNA09DA80A). This work was supported in part by the Jet Propulsion Laboratory, California Institute of Technology, under a contract with NASA. We are grateful to an anonymous referee for helpful comments.

\clearpage


\begin{deluxetable}{cccccc} 
\tabletypesize{\scriptsize} 
\tablecaption{Extinction values and column densities for program stars$^{a}$.}
\tablewidth{0pt} 
\tablehead{ 
\colhead{~$n$} & \colhead{Identification} & \colhead{~~~$A_V$~~~} & \colhead{~$N$(CO)$_{\rm gas}$} & \colhead{~$N$(CO)$_{\rm ice}$} & \colhead{~$N$(CO$_2$)$_{\rm ice}$}\cr
[1]&[2]&[3]&[4]&[5]&[6]}
\startdata 
~1	& J04082673+2803429	& 6.4	& 4.91	& 0.00	& 0.53 \\
~2	& J04090144+2453214	& 4.4	& 3.14	& 0.00	& 0.03 \\
~3	& J04092063+2816031	& 4.1	& 5.10	& 0.00	& 0.00 \\
~4	& J04104300+2820340	& 4.6	& 1.99	& 0.00	& 0.08 \\
~5	& J04110488+2443185	& 3.5	& 2.30	& 0.00	& 0.00 \\
~6	& J04112677+2831093	& 5.5	& 7.67	& 0.00	& 0.30 \\
~7	& J04112801+2830271	& 3.7	& 6.91	& 0.00	& 0.00 \\
~8	& J04113168+2829562	& 4.5	& 6.69	& 0.00	& 0.05 \\
~9	& J04114185+2841282	& 5.9	& 2.51	& 0.00	& 0.40 \\
10	& J04114938+2815568	& 3.6	& 4.92	& 0.00	& 0.00 \\
11	& J04122296+2949441	& 5.6	& 2.85	& 0.00	& 0.33 \\
12	& J04123318+2945152	& 4.8	& 4.75	& 0.00	& 0.13 \\
13	& J04123940+2816419	& 7.0	& 6.64	& 0.12	& 0.68 \\
14	& J04130664+2235365	& 5.2	& 1.88	& 0.00	& 0.23 \\
15	& J04132688+2804584	& 6.3	& 3.52	& 0.00	& 0.50 \\
\enddata
\tablenotetext{a}{The complete table is available in the online version of this paper.\\
Description of columns:
[1]~Running number. [2]~Source identification: field stars from Shenoy \etal\ 2008 \hbox{($n = 1$\,--\,247)} designated by 2MASS catalog number; additional field stars ($n = 248$\,--\,292) mostly designated by Henry Draper (HD) catalog number (Whittet \etal\ 2001); YSOs ($n = 293$\,--\,302) designated by 2MASS catalog number. [3]~Visual extinction in magnitudes. [4]~Gas phase CO column density from the present work in units of $10^{17}\,{\rm cm}^{-2}$. [5] and [6]~Solid phase CO and CO$_2$ column densities, respectively, in units of $10^{17}\,{\rm cm}^{-2}$; direct measurements are included where available (Whittet \etal\ 2007, their table~2, for field stars; Cook et al 2010, their table~3 for YSOs), all other values are calculated using the method described in Section~3.2.}
\end{deluxetable}
\clearpage

\begin{deluxetable}{lcccc} 
\tabletypesize{\scriptsize} 
\tablecaption{Mean and differential CO depletion factors for field stars.}
\tablewidth{0pt} 
\tablehead{ 
\colhead{$A_V^{~a}$} & \colhead{$N$(CO)$_{\rm gas}$} & \colhead{$N$(CO)$_{\rm total}$} & \colhead{$\delta$(CO)$^b$}& \colhead{$\delta$(CO)$^c$}\cr
\colhead{(mag)} & \colhead{($\times 10^{17} {\rm cm}^{-2}$)} & \colhead{($\times 10^{17} {\rm cm}^{-2}$)} & \colhead{(mean)} & \colhead{(difference)}}
\startdata 
~~5   &  3.9  &  4.2  & 0.07 & ---  \\
~10   &  7.0  &  9.6  & 0.27 & 0.43 \\
~20   & 10.2~ & 19.9~ & 0.49 & 0.69 \\
~28.7 & 10.9~ & 25.8~ & 0.58 & 0.88 \\

\enddata
\tablenotetext{a}{Data at $\av = 5$, 10 and 20~mag are averages for typical lines of sight falling within 20\% of these extinction values; data at $\av = 28.7$~mag refer to the individual star J04135352+2813056.}
\tablenotetext{b}{Values calculated (Equation~6) from line-of-sight column densities.}
\tablenotetext{c}{Values calculated (Equation~6) from incremental differences in column densities (current row relative to previous row).}
\end{deluxetable}
\clearpage


\begin{figure}
\centering
\includegraphics[width=15cm, angle=0]{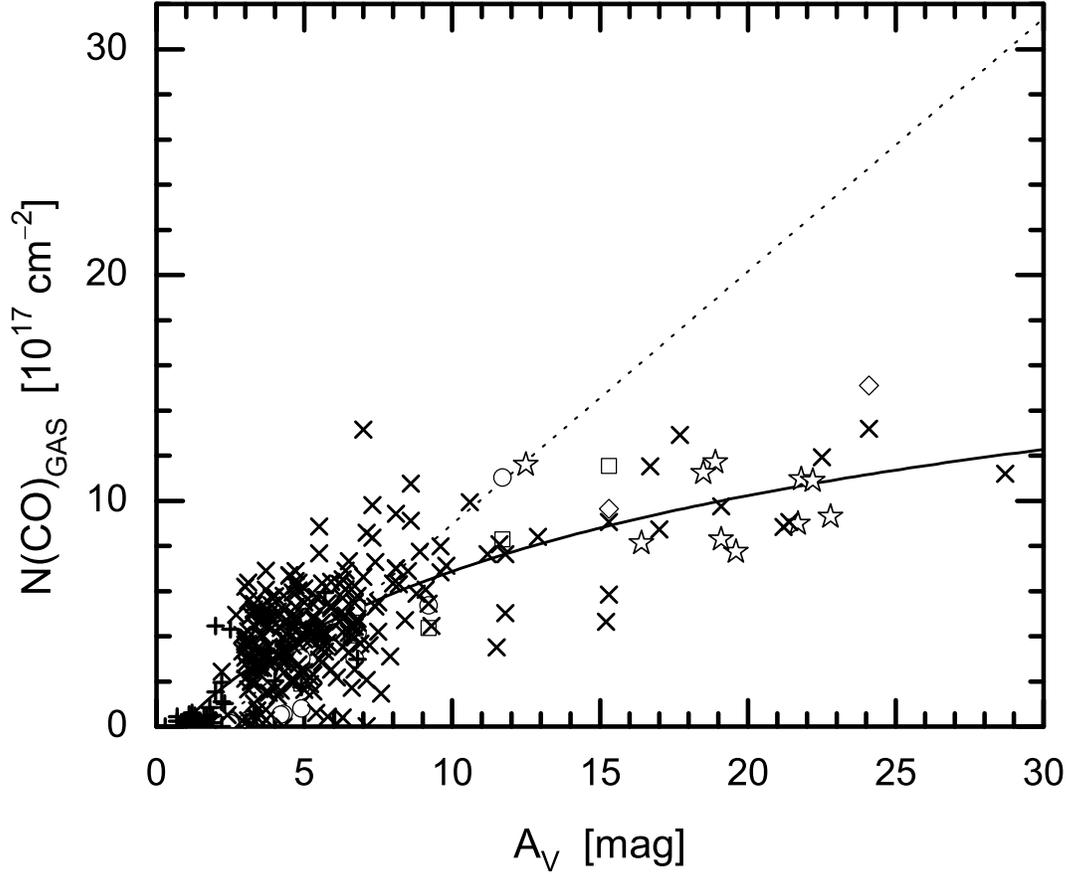}
\caption{\small Plot of gaseous CO column density against visual extinction for lines of sight in our sample (Table~1). Crosses ($\times$) and star symbols represent background field stars and young stellar objects selected from the 2MASS Point Source Catalog (Shenoy \etal\ 2008), respectively. Pluses ($+$) near the origin represent additional optically-selected field stars (Whittet \etal\ 2001). Also plotted are data based on gas-phase $^{13}{\rm C}^{18}$O, C$^{17}$O and C$^{18}$O observations toward selected field stars by Frerking \etal\ (1982), represented by diamonds, squares and circles, respectively. The diagonal dotted line represents the empirical relationship $N({\rm C^{18}O}) = 2 \times 10^{14} (A_V-2)~{\rm cm}^{-2}$ based on data for several molecular clouds (Kainulainen \etal\ 2006 and references therein), assuming $N({\rm CO})/N({\rm C^{18}O})=560$. The solid curve is a sigmoidal fit to the data for the entire sample (see Section~3.1; the fit parameters are $a_1=-1.532$, $a_2=23.202$, $x_0=22.541$ and $p=0.8164$).
 \label{fig1}}
\end{figure}
\clearpage


\begin{figure}
\centering
\includegraphics[width=15cm, angle=0]{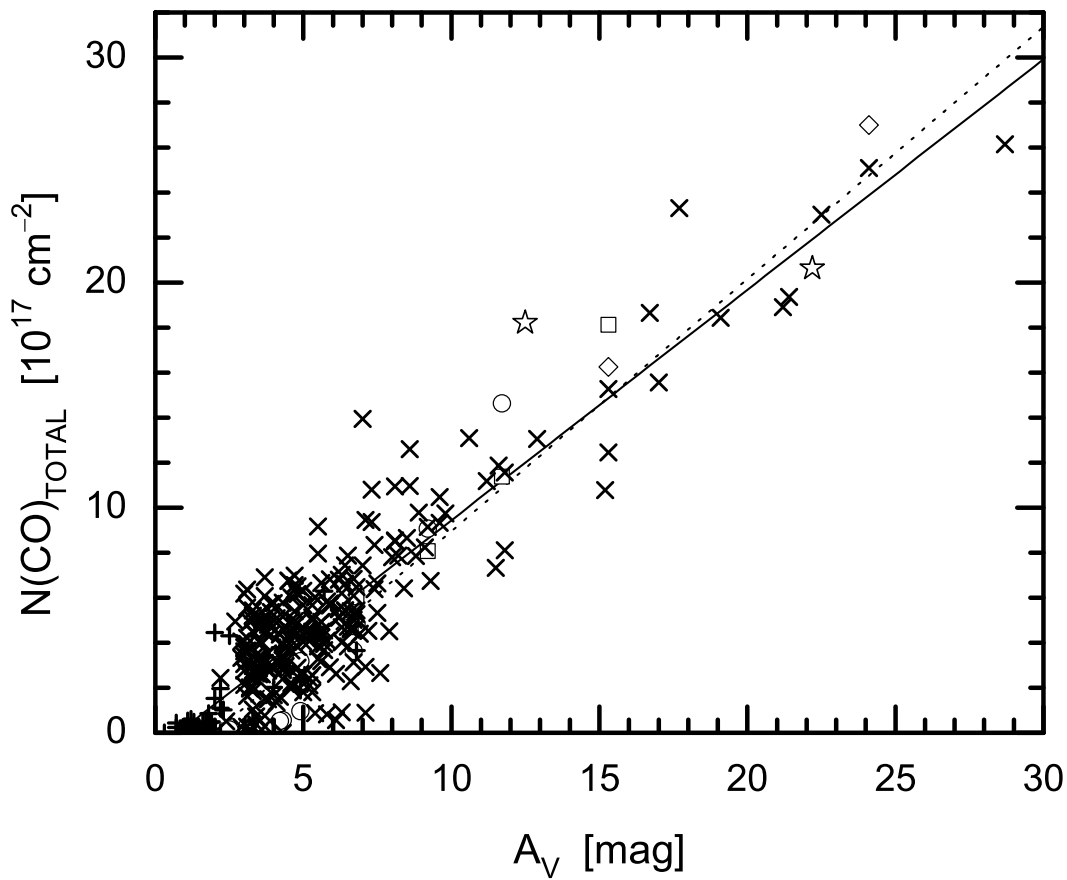}
\caption{\small Plot of total (gas + ice) CO column density against visual extinction. Solid \cod\ is also included in the ice component (see text, section~3.2). Symbols have the same meaning as in \fig 1. The dotted line represents the empirical relationship $N({\rm C^{18}O}) = 2 \times 10^{14} (A_V-2)~{\rm cm}^{-2}$, as in \fig 1. The solid line is the unweighted linear least-squares fit to the data for the entire sample, given by $N({\rm CO})_{\rm total} = (1.024\pm 0.025)[\av - (0.78\pm 0.15)]\times 10^{17}~{\rm cm}^{-2}$.\label{fig2}}
\end{figure}
\clearpage


\begin{figure}
\centering
\includegraphics[width=15cm, angle=0]{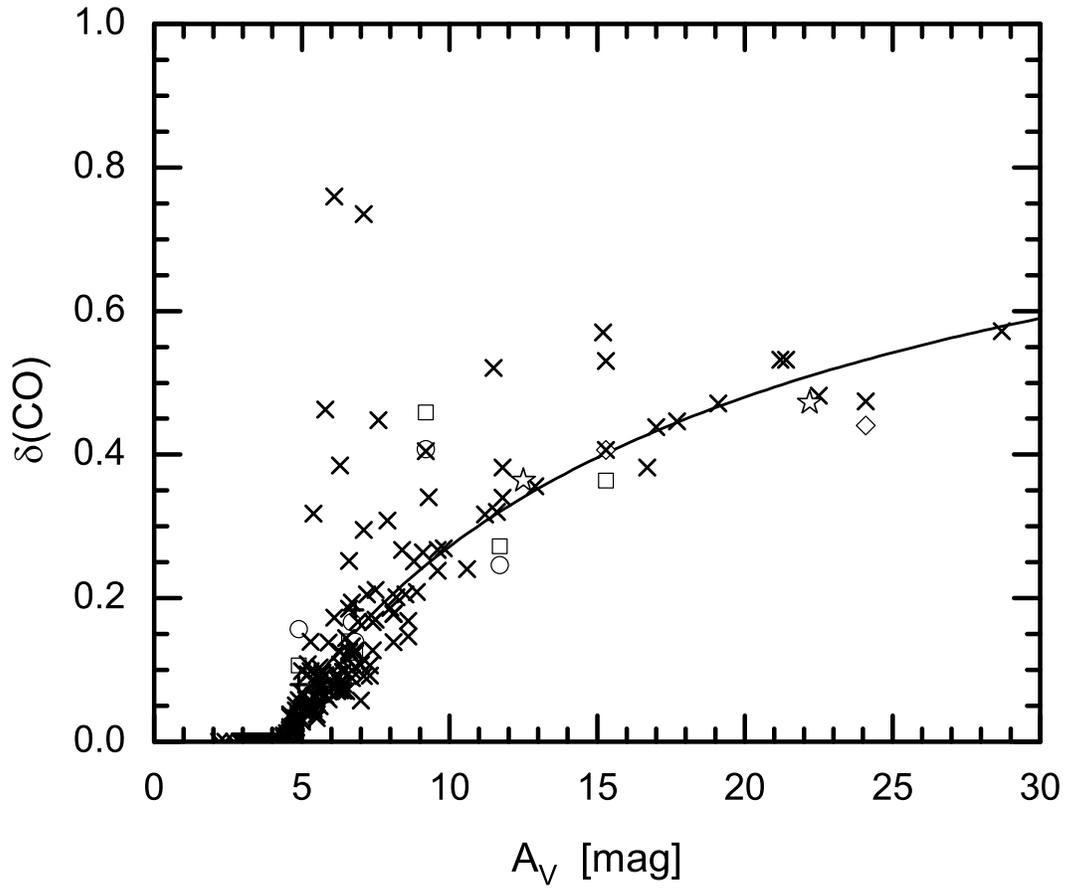}
\caption{\small Plot of mean CO depletion factor $\delta({\rm CO})$, defined in Equation~(6), against visual extinction. Symbols have the same meaning as in Figures~1 and 2. The solid curve is obtained by combining the sigmoidal fit from \fig 1 with the linear fit from \fig 2. \label{fig3}}
\end{figure}
\clearpage

\end{document}